# Observing Molecular Spinning via the Rotational Doppler Effect

Omer Korech,[1] Uri Steinitz,[1] Robert J. Gordon,[2] Ilya Sh. Averbukh,[1] and Yehiam Prior[1]

[1]Department of Chemical Physics, Weizmann Institute of Science, Rehovot 76100, Israel
[2]Department of Chemistry, University of Illinois at Chicago (mc 111), Chicago, IL 60607, United States

When a wave is reflected from a moving object, its frequency is shifted due to the well-known Doppler effect [1,2]. Similarly, when circularly polarized (CP) light is scattered from a rotating object, a rotational Doppler frequency shift (RDS) may be observed, with manifestations ranging from the quantum world (fluorescence spectroscopy, rotational Raman scattering, etc. [3,4]) to satellite-based GPS navigation systems [5]. (For a recent overview of the RDS effect see Ref. 6). In our work, we observe for the first time the RDS phenomenon for a CP light wave propagating through a gas of molecules (nitrogen or deuterium) synchronously spinning in the same sense (clockwise or counter-clockwise). Such a state of the gas medium was achieved by a double-pulse laser excitation, in which the first pulse aligns molecules along a certain direction, and the second one (linearly polarized at a $45^0$ angle) causes a concerted unidirectional rotation of the "molecular propellers" [7,8]. We observed the resulting rotating birefringence of the gas by detecting a Doppler-shifted CP wave of the opposite handedness at the output of the medium. The observed RDS is in the THz range, six orders of magnitude greater than previously measured in mechanically rotating birefringent crystals.

In his famous 1905 paper on special relativity[1], Einstein considered reflection of linearly polarized light of frequency ω from a mirror moving with velocity $v$ and showed that the reflected wave has a frequency $\omega \pm 2kv$ (in the non-relativistic limit), where $k = \omega/c$, $c$ is the speed of light, and the sign depends on the relative direction of the incident wave and the moving mirror. When an anisotropic polarizable object rotates with the angular velocity $\Omega$, a rotational Doppler frequency shift (RDS) may be observed in the scattering of a circularly polarized electromagnetic wave. The scattered field consists of a circularly polarized component having the same frequency and handedness as the incident one and a circularly polarized wave of the opposite handedness with a displaced frequency of $\omega \pm 2\Omega$. This frequency shift results from the exchange of angular momentum and energy between the electromagnetic field and the rotating object [9-13]. Until now, table-top observations of this phenomenon utilized mechanical rotation of anisotropic optical elements [10,11,14], and electro-optic effects in a nonlinear crystal subject to a rotating microwave frequency electric field [15]. In the present study, we use a sequence of ultrashort laser pulses to induce gas phase molecules to rotate rapidly in a predefined direction. The biased molecular rotation results in a rotating birefringence of the gas and induces a rotational Doppler frequency shift in the spectrum of a delayed circularly polarized probe pulse. This spectral shift, measured under field-free conditions, lies in the THz range and is many orders of magnitude larger than that observed in mechanically rotated systems.

We induce unidirectional rotation (UDR) of diatomic molecules by a "molecular propeller" method proposed by Fleischer et al. [7], which uses two short time-delayed non-resonant laser pulses with different linear polarization angles. The first laser pulse induces transient field-free alignment of molecules along the pulse polarization, producing an anisotropic angular distribution. (See Refs. 16 and 17 for a recent review of laser molecular alignment and [18] for earlier studies.) At the time of maximal alignment, a second pulse is applied (see Fig. 1) with its polarization axis set at 45° with respect to that of the first pulse, causing the transiently-aligned molecules to rotate towards the second axis. This technique was demonstrated experimentally by Kitano et al. [8] and subsequently enhanced by Zhdanovich et al. [19], who used a long train of chiral laser pulses to achieve unidirectional rotation (see also Refs. 20 and 21).

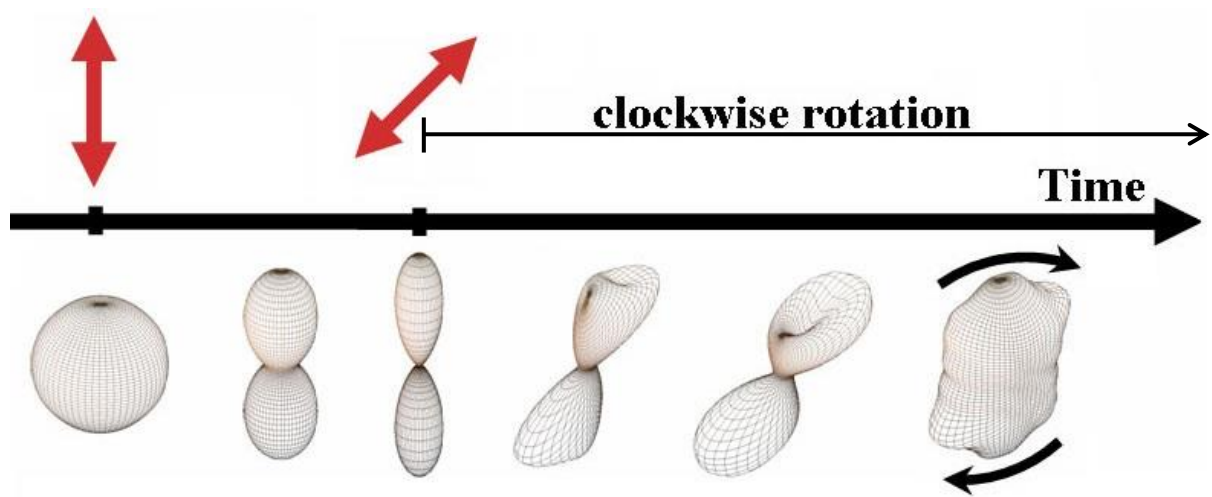

**Figure 1, UDR excitation scheme** (reproduced with permission from [7]): The double arrows indicate the polarizations directions and timing of the pump pulses. The bottom drawings illustrate the angular distribution of the rotating ensemble along the same time axis.

While UDR persists as long as the molecules do not collide, the field-free anisotropy of the angular distribution gradually disappears because of dispersion of the molecular angular velocity. The cigar-like shape of the angular distribution reappears periodically because of quantum revival of the rotational wave packets (see, e.g. Ref. 18) with a revival period of $T_{rev} = 1/(2Bc)$, where $B$ is the molecular rotational constant. Substantial anisotropy of the angular distribution is observed also at fractions of $T_{rev}$, especially near $T_{rev}/2$. Although the most natural time to observe UDR is immediately after the second pump pulse, residual light from that pulse may interfere with the measurement [6,22]. This problem may be overcome by observing the UDR at a full or fractional revival under completely field-free conditions.

Let us now consider in more detail the interaction of a circularly polarized time-delayed probe pulse with an ensemble of unidirectionally rotating molecules. As mentioned previously, the polarizability anisotropy of the aligned molecules produces a significant birefringence. This birefringence is strongest near the time of the first maximal alignment (the cigar-like shape in Fig. 1), or at full and half revival repetitions. The birefringent medium converts a part of the incident electromagnetic field into one with opposite polarization handedness. Moreover, the time varying alignment and birefringence also induce a transient variation in the refractive index . Light propagating through a medium with a time-dependent refractive index experiences phase changes, which broaden and shift its spectrum. This phenomenon, known as Molecular Phase Modulation (MPM), has been used previously for compression and spectral broadening of ultrashort laser pulses [23-25]. Finally, because of the unidirectional rotation of the aligned molecules, not only is the magnitude of the birefringence time-dependent, but also the axes of the polarizability tensor rotate with time. Each element of the gas volume therefore resembles a rotating anisotropic dielectric body, producing a rotational Doppler shift of the frequency of the scattered probe beam, as reported in this paper.

Both MPM and RDS lead to spectral changes of the probe pulse resulting from the non-stationary character of the refraction index. These two effects respond differently, however, to changes in the time delay and polarization of the probe. The MPM spectral shift reverses its sign when the probe delay scans across the alignment (or anti-alignment) maximum. In contrast, the RDS induced by unidirectional molecular rotation is a one-sided spectral shift (red or blue), depending on whether the molecules and the circularly polarized probe rotate in the same or opposite directions. Since the birefringence is greatest at the time of maximum alignment, scanning the delay of the probe pulse around that time provides a means of distinguishing between MPM and RDS. A detailed theory treating the interplay between these two effects will be published elsewhere (U.S., Y.P. and I.S.A, in preparation).

The experimental setup is depicted in Fig. 2. A regeneratively amplified Ti:Sapphire laser (peak wavelength ~790 nm, 50 fsec FWHM) was used to generate the pump and probe pulses. The intensity of the pumps was set to ~$10^{14}$ W/cm$^2$, while the probe intensity was two orders of magnitude lower. The three pulses intersected at the focus of an achromatic lens (f = 100 cm) in a three dimensional boxCARS configuration [26]. The four-wave mixing (FWM) beam generated in the intersection region was used to optimize the spatial and temporal overlap of all the pulses. Both pump pulses were linearly polarized, their relative polarization angle was set by achromatic half wave plates, and the delay between them was set to match the time of maximal molecular alignment.

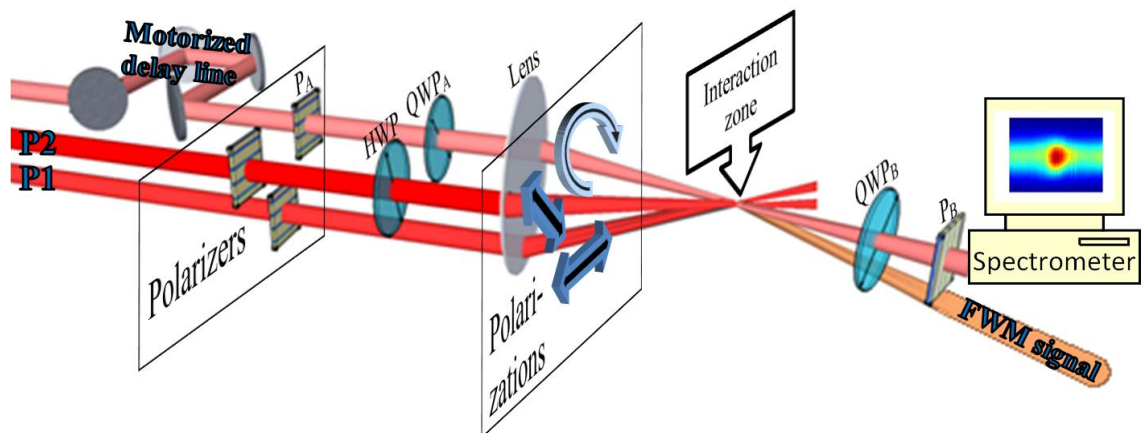

**Figure 2, experimental setup**: Pump P1 is polarized and focused by the lens into the interaction zone. The polarization axis of pump P2 is controlled by the half wave plate, HWP. The probe is circularly polarized by polarizer $P_A$ and quarter wave plate $QWP_A$. The polarizations are indicated by the arrows. P2 delay with respect to P1 is fixed. The delay between the pump and probe pulses is scanned by the motorized delay line. The analyser, consisting of quarter wave plate $QWP_B$ and polarizer $P_B$, is aligned to block the probe in the absence of the pump pulses.

A circularly polarized (CP) probe pulse was introduced after a variable time delay, immediately following the pump pulses or near the half revival time. A circular analyser (composed of a quarter wave plate and a linear polarizer) intercepting the probe trajectory between the interaction region and the spectrometer was aligned to block the probe in the absence of the pump pulses. The RDS light generated by the unidirectionally rotating molecules has a reverse handedness, allowing it to pass through the analyser. The spectrum of the transmitted probe measured at different delays yielded a spectrogram, from which the time evolution of the frequency shift was determined.

The first set of measurements was performed using deuterium molecules (D$_2$) contained in a gas cell. For this light molecule, the induced UDR is faster, and therefore the expected Rotational Doppler shift is larger than that expected for nitrogen. Figure 3 displays three representative spectrograms, in which the transmitted probe spectrum is plotted in the XZ plane for various probe delays (Y-axis) in a three-dimensional plot. In each panel, the black line indicates the center frequency of the incident probe. The spectrogram in panel a) is the result of a control experiment, where only one pump pulse was introduced, producing a transient alignment (but no UDR). The spectrogram in panels b) and c) were recorded when the molecules were set to rotate in the same and opposite sense as the CP probe, respectively, resulting in red and blue shifts of the transmitted light. Panels b,c display sizable unidirectionally-induced rotational Doppler shifts of the probe that are properly correlated with the sense of the induced molecular rotation.

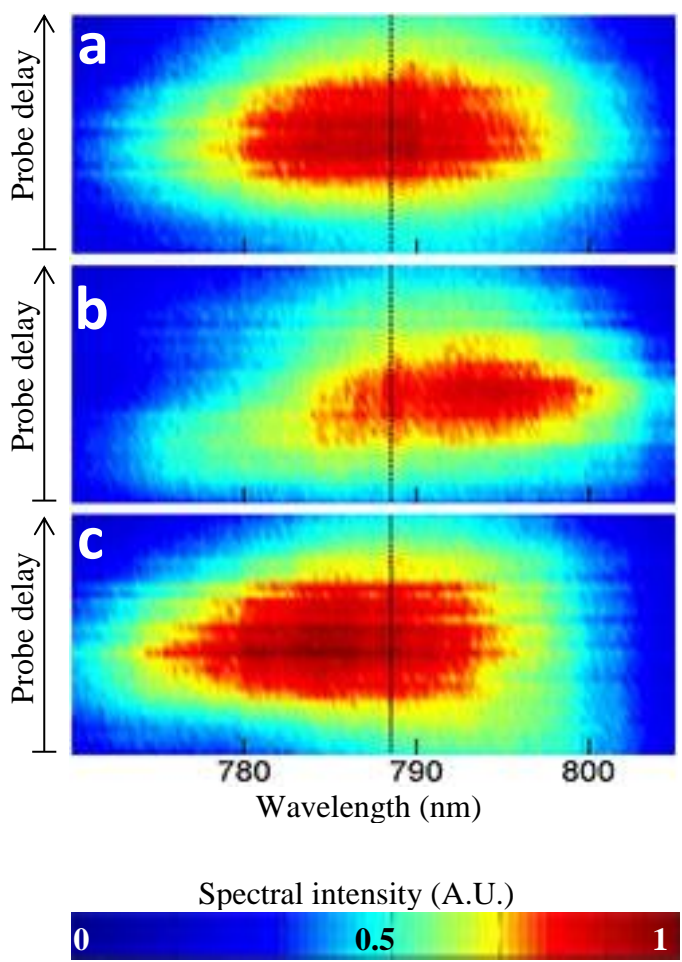

**Figure 3, Measured rotational Doppler shift for deuterium.** The transmitted probe spectrum is plotted against the probe delay following the pump pulses. In each panel the probe scans a delay of 150 fsec around the peak. a) Only one pump is applied, resulting in no UDR and no RDS. b) The molecules are set to rotate in the same sense as the CP probe, producing red shift; c) The molecules are set to rotate in opposite sense, producing blue shift. The spectra are recorded in 5 fsec intervals. The dotted black line marks the central frequency of the unperturbed probe.

The next set of experiments was performed using nitrogen molecules under ambient conditions. These experiments were conducted in the open atmosphere avoiding any birefringence originating in the sample cell windows. The presence of atmospheric oxygen does not pose a problem; the larger moment of inertia of O$_2$ gives it a longer revival time

(11.9 psec as compared to 8.3 psec for $N_2$) so that its angular distribution is essentially isotropic around the time of the $N_2$ revival, when the measurement is performed.

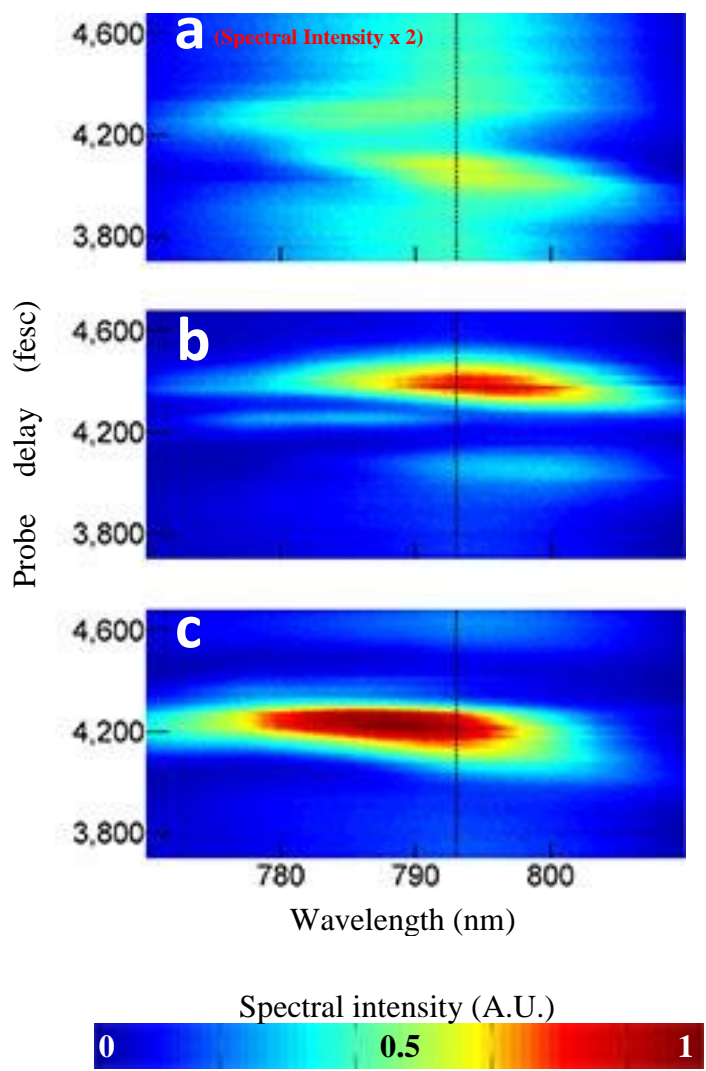

**Figure 4, Measured rotational Doppler shift for nitrogen molecules.** In each panel the transmitted probe spectrum is plotted against the probe delay around the half revival time. a) Only one pump is applied, resulting in no UDR. The signature of MPM is observed. Note that the intensity in this panel was multiplied by a factor of 2 to enhance visibility. b) The molecules are set to rotate in the same sense as the CP probe, producing a red shift. c) The molecules are set to rotate in the opposite sense of the probe, producing a blue shift. The spectra are recorded in 10 fsec steps. The dotted line indicates the unperturbed probe central frequency.

Figure 4 depicts the RDS near the half revival time of $N_2$. As in Fig. 3, panel a) corresponds to the control experiment, where only one pump pulse is present, resulting in a transient alignment (but no UDR). In this panel, evidence for MPM is clearly observed. The MPM spectral shift changes its sign when the delay crosses the region of the maximal signal, as discussed above and reported earlier [23-25]. Note, that the intensity in this panel was multiplied by a factor of 2 to enhance visibility. The spectrogram in panel b) was recorded when the molecules were set to rotate in the same sense as the CP probe, resulting in a red shift of a few nanometers to longer wavelengths. When the molecules were set to rotate in the opposite direction (by readjusting the polarization direction of the second pump), a blue-shifted spectrum is evident (panel c). The observed rotational Doppler shifts are properly correlated with the sense of the induced molecular rotation.

We performed several other control experiments (not shown here) to confirm our observation of the RDS caused by unidirectional molecular rotation. In particular, instead of changing the angle between the polarization axes of the two pumps with a half wave plate, we reversed the direction of the UDR by switching the time delay between the pumps. In another experiment, a notch filter was introduced to imprint a narrow spectral feature on the probe spectrum, and a shift of this feature was measured as a function of the probe delay. In all cases, a clear unidirectional RDS was recorded that was correlated with the sense of molecular rotation.

In conclusion, we used coherently spinning molecules to generate a rotational Doppler shift that is six orders of magnitude greater than previously observed. To achieve this result we used a linearly polarized pulsed laser beam to align a thermal ensemble of diatomic molecules followed by a second linearly polarized pulse to kick the molecular axes in a certain direction. The resulting unidirectionally rotating molecules were probed by a circularly polarized beam of light. Upon transmission through the spinning molecules, a part of the probe light inverted its handedness and acquired a red or blue spectral shift, depending on whether the molecules were co- or counter-rotating relative to the probe sense of circular polarization. This experiment provides direct evidence of unidirectional rotation of an ensemble of molecules and paves the way for a new class of measurements in which the rotational direction of a molecular reagents may be monitored or actively controlled.

Financial support of this research by the Israel Science Foundation (Grant No. 601/10) and the Deutsche Forschungsgemeinschaft (Grant No. LE 2138/2-1) is gratefully acknowledged. IA is the incumbent of the Patricia Elman Bildner Professorial Chair. YP is the incumbent of the Sherman Professorial Chair. This research is made possible in part by the historic generosity of the Harold Perlman Family. RJG thanks the Weston foundation for support during a sabbatical visit.